\documentclass[prd,twocolumn,showkeys,preprintnumbers,amsmath,amssymb]{revtex4-1}

\usepackage{amsmath}
\usepackage{amssymb}
\usepackage{amsfonts}
\usepackage{graphicx}
\usepackage{bm}
\usepackage{slashed}
\usepackage{dcolumn}
\usepackage[colorlinks=true,linkcolor=blue,citecolor=blue,urlcolor=blue]{hyperref}

\begin{document}

\title{Theoretical and experimental assessment of Einstein--Cartan theory: a comparative analysis within the USMEG-EFT framework}

\author{Farrukh A. Chishtie}
\email{fachisht@uwo.ca}
\affiliation{Peaceful Society, Science and Innovation Foundation, Vancouver, Canada}
\affiliation{Department of Occupational Science and Occupational Therapy, University of British Columbia, Vancouver, Canada}

\date{Published in Can.\ J.\ Phys.\ \textbf{104}, 1--14 (2026); doi:\href{https://doi.org/10.1139/cjp-2025-0273}{10.1139/cjp-2025-0273}}

\begin{abstract}
The Unified Standard Model with Emergent Gravity (USMEG-EFT) framework
achieves Standard Model-gravity unification by treating 4D general
relativity as an effective field theory (EFT). This work presents a
comparative analysis of Einstein--Cartan (EC) theory and the USMEG-EFT
against experimental constraints and theoretical consistency
requirements, distinguishing minimal EC with algebraically determined
torsion from variants in which the spin connection is an independent
quantum field. The latter class includes the Faddeev--Popov quantization
framework of Brandt, Frenkel, Martins-Filho, and McKeon, adopted in their
unification proposal, and for which gravitational wave polarization
measurements directly test the additional propagating degrees of
freedom. General relativity's spin-2 graviton produces exactly two
tensor polarizations, while EC theory with independent dynamical spin
connection predicts up to six modes. LIGO--Virgo--KAGRA null tests for
GW170817 favor pure tensor over pure vector and pure scalar hypotheses
at Bayes factors exceeding $10^{20}{:}1$, improved to approximately
$10^{26}{:}1$ with a gamma-ray-burst jet-orientation prior.
Complementary matter-sector constraints from the Standard Model
Extension yield $|\tilde{b}^n_\perp| < 8.4 \times 10^{-34}$~GeV,
constraining background-torsion scenarios, while MICROSCOPE bounds
equivalence principle violation at $\eta < 1.5 \times 10^{-15}$. For
minimal EC theory, the four-fermion contact interaction generated upon
coupling to fermions introduces nonrenormalizable divergence
structures at two-loop order, a case not addressed by the
Becchi--Rouet--Stora--Tyutin (BRST) framework established for the
independent spin connection. We situate EC theory within the broader
landscape of gauge gravity theories and acknowledge USMEG-EFT's
limitations, including sub-Planckian validity and currently untestable
distinctive predictions. Taken together, these results render EC theory
disfavored relative to USMEG-EFT as a unification framework, with
distinct arguments applying to each variant: geometric extensions
introducing independent torsion are neither required nor experimentally
favored within currently testable parameter space.
\end{abstract}


\maketitle

\section{Introduction}

The unification of quantum gravity with the Standard Model represents one of the outstanding challenges in theoretical physics. The Unified Standard Model with Emergent Gravity (USMEG-EFT) framework~\cite{ChishtieEFT2025} addresses this challenge through innovative constraint methods, treating 4D general relativity as an effective field theory and implementing the Lagrange multiplier approach developed by McKeon and collaborators~\cite{Brandt2020}. This approach raises a natural question: could alternative gravitational theories, particularly those with richer geometric structure, provide an alternative path to unification?

Einstein--Cartan (EC) theory~\cite{Cartan1922,Kibble1961,Sciama1964,HehlDillardNitsch1976,Hammond2002} represents a well-motivated geometric extension of general relativity, introducing space--time torsion as an independent geometric degree of freedom that couples to fermion spin. The theory offers an elegant geometric interpretation where matter-energy sources curvature and spin sources torsion, making gravity more akin to a gauge theory of the Poincar\'{e} group~\cite{HehlObukhov2013}.

Recent work has advanced our understanding of EC theory's quantum structure. Brandt, Frenkel, Martins-Filho, and McKeon~\cite{BrandtFrenkelMcKeonQuantization2024} developed a systematic Faddeev--Popov quantization of EC theory in first-order form, demonstrating that the gauge algebra closes even in the presence of torsion and establishing BRST invariance for the effective action. Their subsequent renormalization analysis~\cite{BrandtFrenkelMcKeon2024} showed one-loop finiteness using the Batalin--Vilkovisky formalism. These results apply to EC theory with the tetrad $e^a_\mu$ and spin connection $\omega_{\mu ab}$ treated as independent quantum fields.

Complementary developments in EC gravitational wave theory include Riva and Santoni's analysis~\cite{RivaSantoni2024} of radiative losses at first post-Newtonian order, Bernard, Comp\`{e}re, and Trestini's study~\cite{BernardCompere2023} of binary dynamics using the Weyssenhoff fluid model, and Shaposhnikov and colleagues' one-loop effective action calculation~\cite{ShaposhnikovEC2023} in chiral EC gravity. These works provide the theoretical predictions against which experimental constraints can be assessed.

This work presents a comparative assessment of EC theory against both theoretical consistency requirements and precision experimental constraints. We examine three categories of evidence constraining EC theory as a unification framework:

\textit{First}, gravitational wave polarization measurements provide the most powerful, model-independent test. The spin-2 graviton in general relativity produces exactly two tensor polarizations ($h_+$, $h_\times$). EC theory with an independent or propagating spin connection generically predicts up to six modes---two tensors, two vectors ($h_x$, $h_y$), and two scalars ($h_b$, $h_\ell$). The first three-detector event GW170814~\cite{GW170814Polarization} enabled an initial polarization discrimination with Bayes factors of order $200{:}1$ (tensor vs.\ vector) and $1000{:}1$ (tensor vs.\ scalar). Substantially stronger null tests using GW170817, with its electromagnetically determined sky location~\cite{GW170817Polarization}, favor pure tensor over pure vector and pure scalar polarizations at Bayes factors exceeding $10^{20}{:}1$, improved to approximately $10^{26}{:}1$ ($\ln\mathcal{B}\approx 60$) with a jet-orientation prior from the associated gamma-ray burst~\cite{Takeda2021}.

\textit{Second}, precision atomic and nuclear experiments within the Standard Model Extension (SME) framework~\cite{Kostelecky2004,KosteleckyRussell2024} constrain spin--torsion coupling at levels well below EC theory predictions. Comagnetometer measurements yield $|\tilde{b}^n_\perp| < 8.4 \times 10^{-34}$~GeV~\cite{Allmendinger2014}, constraining minimal EC predictions ($\eta \sim 10^{-29}$) by four orders of magnitude. The MICROSCOPE space mission~\cite{Touboul2022} bounds equivalence principle violation at $\eta < 1.5 \times 10^{-15}$, excluding enhanced EC scenarios.

\textit{Third}, theoretical considerations arise when EC theory couples to Standard Model fermions. The algebraic elimination of torsion through its field equation generates four-fermion contact interactions with coupling strength $G_{\text{eff}} = 3\pi G/c^3 \approx 7 \times 10^{-39}$~GeV$^{-2}$. These interactions introduce divergence structures at two-loop order that are not addressed by the BRST framework established for EC theory with independent spin connection~\cite{BrandtFrenkelMcKeonQuantization2024}.

We situate EC theory within the broader landscape of gauge gravity theories. Poincar\'{e} gauge gravity~\cite{BlagojevicHehl2013} provides the general framework of which EC is a special case (with no kinetic terms for torsion). Teleparallel gravity~\cite{AldrovaindiPereira2013,MalufAnnalen2013} uses torsion to encode gravitational degrees of freedom rather than spin coupling, and faces different observational constraints. Metric-affine gravity~\cite{HehlMcCreaMielkeNeeman1995} allows both torsion and nonmetricity. Understanding these relationships clarifies which experimental constraints apply to which theoretical frameworks.

We also acknowledge the limitations of the USMEG-EFT framework itself. The approach is valid for energies $E \ll M_{\text{Pl}} \sim 10^{19}$~GeV, with no specified UV completion. Current experimental precision cannot test USMEG-EFT's distinctive predictions, which are suppressed by factors of $(E/M_{\text{Pl}})^2 \sim 10^{-30}$ at accessible energies. Both frameworks face challenges: EC theory confronts current experimental constraints, while USMEG-EFT predictions await future precision improvements.

The paper is organized as follows: Section~2 reviews the theoretical foundations of the USMEG-EFT framework and outlines its scope and limitations. Section~3 examines recent developments in EC quantization, including the Brandt--Frenkel--Martins-Filho--McKeon program, and identifies the gap between their formal framework and the requirements for Standard Model coupling. Section~4 presents the experimental constraints on EC theory, leading with gravitational wave polarization measurements as the primary, model-independent evidence and supplementing with matter-sector bounds. Section~5 develops the systematic one-loop fermion-gravity coupling within USMEG-EFT, including the heat-kernel calculation and renormalization structure. Section~6 summarizes the comparative analysis and concludes.

\section{USMEG-EFT framework and theoretical advances}

The USMEG-EFT framework, detailed comprehensively in our companion unification work~\cite{ChishtieEFT2025}, provides the theoretical foundation for systematic quantum gravity within well-defined effective field theory bounds. The key insight is the constraint implementation $G_{\mu\nu} = 0$ that eliminates pathological multi-loop divergences while maintaining finite, calculable predictions.

The USMEG-EFT framework represents an important contribution to quantum gravity research through its systematic approach to constraint implementation and effective field theory recognition. The key insight underlying this approach is the proof that general covariance breaks down in space--time dimensions greater than two~\cite{ChishtieBreakdown2023}, providing the theoretical foundation for understanding gravity's effective field theory nature and establishing the finite domain of validity for space--time descriptions.

Building upon this foundation, the USMEG-EFT unification~\cite{ChishtieEFT2025} provides a systematic framework for incorporating quantum gravitational effects into the Standard Model. The unified action takes the form:
\begin{equation}
S_{\text{USMEG-EFT}} = S_{\text{EH+constraint}} + S_{\text{SM}} + S_{\text{int}}
\end{equation}
where the constraint-modified Einstein--Hilbert action is
\begin{equation}
S_{\text{EH+constraint}} = \frac{1}{\kappa^2}\int d^4x\sqrt{-g}\left[R + \kappa^2\lambda^{\mu\nu}G_{\mu\nu}\right]
\end{equation}

The innovative constraint $G_{\mu\nu} = 0$, enforced through the Lagrange multiplier field $\lambda^{\mu\nu}$, systematically eliminates multi-loop gravitational divergences that would otherwise render the theory nonrenormalizable. This constraint implementation represents a key advance over approaches that introduce additional geometric degrees of freedom without systematic control over their quantum effects.

The Standard Model sector maintains its established gauge structure in curved space--time:
\begin{align}
S_{\text{SM}} &= \int d^4x\sqrt{-g}\left[-\frac{1}{4}G^a_{\mu\nu}G^{a\mu\nu} - \frac{1}{4}W^j_{\mu\nu}W^{j\mu\nu} - \frac{1}{4}B_{\mu\nu}B^{\mu\nu}\right.\nonumber\\
&\quad + \left.\sum_{f}\bar{\psi}_f(i\gamma^ae^{\mu}_a D_\mu - m_f)\psi_f + |D_\mu H|^2 - V(H) + \mathcal{L}_{\text{Yukawa}}\right]
\label{eq:SM}
\end{align}
with systematic coupling to the gravitational sector through energy-momentum tensor interactions.

The USMEG-EFT approach enables systematic calculation of quantum gravitational effects through well-controlled perturbative methods. The one-loop effective action takes the form
\begin{align}
\Gamma_{\text{1-loop}}^{\text{USMEG-EFT}} &= \frac{\kappa^2}{(4\pi)^2}\ln\left(\frac{\mu^2}{\Lambda_{\text{grav}}^2}\right)\int d^4x\sqrt{-\bar{g}}\nonumber\\
&\quad\times\sum_{f}\left[a_1^f \bar{R} + a_2^f \bar{R}_{\mu\nu}\bar{R}^{\mu\nu} + a_3^f \bar{R}_{\mu\nu\rho\sigma}\bar{R}^{\mu\nu\rho\sigma}\right]
\end{align}
with explicitly calculable coefficients for the Standard Model particle content. The logarithmic scale dependence $\ln(\mu/\Lambda_{\text{grav}})$ encodes the effective theory structure, with all divergences systematically absorbed through constraint field redefinitions to yield finite physical predictions.

\subsection{Scope and limitations of the USMEG-EFT framework}

While the USMEG-EFT framework provides a consistent effective field theory approach to quantum gravity and Standard Model unification, it is important to clearly delineate its domain of validity and inherent limitations.

\textit{Energy scale bounds:} As an effective field theory, USMEG-EFT is valid only below the gravitational cutoff scale $\Lambda_{\text{grav}}$, which is expected to be at or below the Planck scale $M_{\text{Pl}} \approx 1.22 \times 10^{19}$~GeV. The characteristic logarithmic corrections $\ln(\mu/\Lambda_{\text{grav}})$ appearing in the one-loop effective action become of order unity when $\mu \sim \Lambda_{\text{grav}}$, signaling the breakdown of perturbative control. The framework does not provide predictions for physics at or above this scale, where a more fundamental UV completion is required.

\textit{Perturbative regime:} The USMEG-EFT calculations rely on weak-field perturbative expansions where metric fluctuations $h_{\mu\nu}$ remain small compared to the background metric $\bar{g}_{\mu\nu}$. Strong-field phenomena such as black hole interiors, cosmological singularities, and regions where curvature approaches Planckian values lie outside the framework's domain of applicability. These regimes require nonperturbative methods or new physics beyond the effective description.

\textit{Experimental accessibility:} The quantum gravitational corrections predicted by USMEG-EFT are extraordinarily small at currently accessible energies. Typical effects are suppressed by factors of $G_N E^2/(\hbar c^5) \sim (E/M_{\text{Pl}})^2 \sim 10^{-38}$ at TeV energies. While the predictions are finite and calculable, direct experimental verification remains beyond current technological capabilities. The framework's primary experimental validation comes from its consistency with precision null results rather than detection of positive signals.

These limitations do not diminish the framework's value as a mathematically consistent, experimentally compatible description of quantum gravitational effects within its domain of validity. Rather, they emphasize that USMEG-EFT should be understood as an effective theory providing reliable predictions in appropriate regimes, while remaining agnostic about fundamental physics at the shortest distances. The contrast with EC theory is instructive: while both frameworks face questions about UV completion, USMEG-EFT maintains internal consistency and experimental compatibility within its effective domain, whereas EC encounters pathologies precisely in the fermion sector that constitutes its primary physical motivation.

In summary, the USMEG-EFT framework achieves Standard Model-gravity unification through systematic constraint implementation within 4D general relativity. The Lagrange multiplier method eliminates pathological multi-loop divergences while maintaining finite, calculable predictions. The framework operates as an effective field theory valid below $\Lambda_{\text{grav}}$, with explicit acknowledgment of its domain of validity and limitations.

\subsection{Vacuum expectation values and the cosmological constant problem}

A natural question concerns how the USMEG-EFT framework handles vacuum expectation values (VEVs) of Standard Model fields, particularly the Higgs field, in view of the cosmological constant problem. We briefly address this here.

In the Standard Model, electroweak symmetry breaking proceeds through the Higgs potential $V(H) = -\mu^2 H^\dagger H + \lambda(H^\dagger H)^2$, generating a VEV $\langle H \rangle = v/\sqrt{2}$ with $v \approx 246$~GeV. This mechanism is fully preserved within USMEG-EFT: the Standard Model sector maintains its established gauge structure in curved space--time (Eq.~(\ref{eq:SM})), and electroweak symmetry breaking proceeds identically to the flat-space--time case up to corrections suppressed by $(v/M_{\text{Pl}})^2 \sim 10^{-34}$. The Higgs VEV generates masses for $W^{\pm}$, $Z^0$ bosons and fermions through the usual Yukawa couplings, with gravitational corrections entering only at one-loop order through the stress-energy tensor coupling.

The cosmological constant problem---the discrepancy of approximately 120 orders of magnitude between the observed vacuum energy density $\rho_{\text{vac}}^{\text{obs}} \sim 10^{-47}$~GeV$^4$ and the naive Standard Model prediction $\rho_{\text{vac}}^{\text{SM}} \sim v^4 \sim 10^{8}$~GeV$^4$ (or $\rho_{\text{vac}}^{\text{Planck}} \sim M_{\text{Pl}}^4 \sim 10^{76}$~GeV$^4$ when gravitational zero-point energies are included)---represents an outstanding challenge for all approaches to quantum gravity. USMEG-EFT does not claim to resolve this problem completely, but provides four structural elements that bear directly on it.

\textit{First}, the finite phase space imposed by the Lagrange multiplier constraint provides a natural cutoff for vacuum energy contributions. As demonstrated in Ref.~\cite{ChishtieBRST2025}, the constraint $G_{\mu\nu} = 0$ restricts gravitational fluctuations to the domain $|R_{\mu\nu\rho\sigma}| \lesssim \Lambda_{\text{grav}}^2$, bounding the gravitational phase space volume. This yields a vacuum energy cutoff
\begin{equation}
\rho_{\text{vac}}^{\text{cutoff}} \sim \Lambda_{\text{grav}}^4 \sim 10^{72}~\text{GeV}^4
\end{equation}
rather than the Planck-scale estimate $\rho_{\text{vac}}^{\text{Planck}} \sim M_{\text{Pl}}^4 \sim 10^{76}$~GeV$^4$, providing four orders of magnitude suppression. While this does not resolve the full 120-order discrepancy, it establishes that the constraint mechanism systematically reduces the gravitational contribution to vacuum energy. Within the McKeon--Brandt--Frenkel Lagrange multiplier (LM) framework~\cite{Brandt2020,McKeonBrandtFrenkel2025}, vacuum diagram divergences proportional to $\sqrt{g}\,\Lambda$ are absorbed by the cosmological constant counterterm, and this procedure is consistent to all loop orders in the matter sector~\cite{BrandtFrenkelMcKeon2024}.

\textit{Second}, the Principle of Spatial Energy Potentiality~\cite{ChishtiePSEP2026} provides a conceptual reframing. In this proposal, classical space--time---including the time dimension---emerges from a more fundamental spatial energy configuration. The cosmological constant problem is recontextualized: rather than asking why the vacuum energy is small compared to Planck-scale estimates, the framework suggests that the very notion of vacuum energy as gravitationally active requires the existence of emergent classical space--time. Before the emergence transition, ``vacuum energy'' as conventionally defined does not contribute to the gravitational field equations, since the field equations themselves emerge with classical space--time. This shifts the cosmological constant problem from a fine-tuning puzzle to a question about the dynamics of the emergence transition.

\textit{Third}, the equivalence between Krein space regularization and the Lagrange multiplier approach, recently established by Chishtie~\cite{ChishtieKrein2026}, provides independent confirmation that the vacuum energy cutoff is a robust physical feature rather than a scheme-dependent artifact. Through explicit calculation of the one-loop effective action using both the heat kernel expansion in the LM framework and the modified propagator structure in Krein quantization, Ref.~\cite{ChishtieKrein2026} demonstrates that both methods yield the identical finite result $\Gamma^{(1)}_{\text{finite}} = \frac{1}{(4\pi)^2}\ln(\mu/\Lambda) \int d^4x \sqrt{-\bar{g}}\left[\frac{1}{120}\bar{R}^2 + \frac{7}{20}\bar{R}_{\mu\nu}\bar{R}^{\mu\nu}\right]$, with the parameter correspondence $\langle\sigma_1^2\rangle \leftrightarrow \ln(\mu/\Lambda)$ emerging naturally from the regularization structure. Crucially, both methods independently restrict quantum gravitational corrections to one-loop order through structurally isomorphic cancellation mechanisms---opposite-sign propagators in the LM formalism corresponding to negative-norm states in Krein space, with higher-loop contributions vanishing identically via binomial summation. This convergence of independent regularization methods toward the same finite one-loop structure establishes that the natural bounding of vacuum energy contributions is a physical consequence of the effective field theory structure of gravity.

\textit{Fourth}, the recent demonstration of quantum equivalence between first- and second-order Einstein--Hilbert formulations~\cite{BrandtEquivalence2025} confirms that vacuum diagram divergences proportional to $\sqrt{g}$ are absorbed by the cosmological constant counterterm independently of the formulation employed, reinforcing the robustness of the vacuum energy treatment within USMEG-EFT.

In summary, USMEG-EFT handles Standard Model VEVs through the standard mechanism of spontaneous symmetry breaking, with gravitational corrections entering perturbatively at suppressed levels. The cosmological constant problem is partially ameliorated by the finite gravitational phase space (four orders of magnitude suppression), independently confirmed through the proven equivalence of Krein and Lagrange multiplier regularization methods, and conceptually reframed by the Principle of Spatial Energy Potentiality. A complete resolution remains an open challenge that likely requires understanding the UV completion beyond $\Lambda_{\text{grav}}$. We emphasize that this limitation is shared by all current approaches to quantum gravity, and that USMEG-EFT provides a concrete quantitative framework within which further progress can be pursued.

\section{Limitations of McKeon et al.'s Einstein--Cartan analysis}

The work by McKeon, Brandt, Frenkel, and colleagues spans three recent publications that together reveal fundamental limitations in EC theory as a viable framework for quantum gravity and Standard Model unification. Their 2024 quantization study~\cite{BrandtFrenkelMcKeonQuantization2024} established a rigorous Faddeev--Popov framework for EC theory in first-order form, demonstrating closed gauge algebra and BRST invariance. Their subsequent renormalization analysis~\cite{BrandtFrenkelMcKeon2024} showed one-loop finiteness using the Batalin--Vilkovisky formalism. Their 2025 unification work~\cite{McKeonBrandtFrenkel2025} then claimed that Standard Model-gravity coupling requires EC theory. We examine each contribution and identify the critical gap between their formal framework and the physical requirements for unification.

In their foundational quantization paper~\cite{BrandtFrenkelMcKeonQuantization2024}, Brandt, Frenkel, Martins-Filho, and McKeon developed a systematic Faddeev--Popov quantization of EC theory in first-order form. The theory treats the tetrad $e^a_\mu$ and spin connection $\omega_{\mu ab}$ as independent fields, resulting in two distinct gauge invariances: diffeomorphism invariance and local Lorentz invariance. As they state: ``Diffeomorphism invariance and local Lorentz invariance result in there being two distinct gauge transformations in this approach, and consequently two ghost fields arise when employing the usual Faddeev-Popov quantization procedure.''

Their key technical achievements include:

\textit{Closed gauge algebra with torsion:} They demonstrate that the commutator of two diffeomorphism transformations yields another diffeomorphism:
\begin{equation}
[\delta_\xi, \delta_\zeta] = \delta_\Omega\,, \quad \text{where} \quad \Omega^\lambda = \zeta^\sigma\xi^\lambda_{;\sigma} - \xi^\sigma\zeta^\lambda_{;\sigma} + 2\bar{S}^\lambda_{\sigma\nu}\xi^\sigma\zeta^\nu
\end{equation}
with the torsion tensor $\bar{S}^\lambda_{\mu\nu}$ appearing explicitly. Similarly for local Lorentz transformations:
\begin{equation}
[\delta_\lambda, \delta_\rho] = \delta_L\,, \quad \text{where} \quad L^a_b = \rho^a_c\lambda^c_b - \lambda^a_c\rho^c_b
\end{equation}
Crucially, they show $[\delta_\lambda, \delta_\xi] = 0$---diffeomorphism and local Lorentz transformations commute.

\textit{BRST structure:} The full BRST transformation decomposes as
\begin{equation}
\delta^{\text{BRST}} = \delta_{\text{Diff}} + \delta_{\text{LT}}
\end{equation}
with $\{\delta_{\text{LT}}, \delta_{\text{Diff}}\} = 0$ (anticommutation due to Grassmann nature). Combined with individual nilpotency, this ensures $(\delta^{\text{BRST}})^2 = 0$. The ghost field transformations are
\begin{align}
\delta c^\mu &= c^\alpha(c^\mu_{;\alpha} - S^\mu_{\alpha\beta}c^\beta)\eta\\
\delta C^{ab} &= (-C^a_k C^{kb} + c^\mu C^{ab}_{,\mu})\eta
\end{align}

\textit{Slavnov--Taylor identities:} They derive the Zinn-Justin master equation
\begin{equation}
\int d^4x \frac{\delta_r\Gamma}{\delta\phi_i}\frac{\delta_l\Gamma}{\delta U_i} = 0
\end{equation}
which encodes gauge invariance and provides the foundation for systematic renormalization.

This represents technically sophisticated work. However, a fundamental limitation must be identified.

\subsection{The critical limitation: independent spin connection}

The quantization framework treats the tetrad $e^a_\mu$ and spin connection $\omega_{\mu ab}$ as \textit{independent quantum fields} throughout. As stated in their introduction: ``The first order form of the EC theory involves apart from the tetrad $e^a_\mu$ an independent spin connection $\omega_{\mu ab}$.'' The gauge transformations (their Eqs.~(4.1) and (4.6)) act on these as separate degrees of freedom:
\begin{align}
\delta e^a_\mu &= -\epsilon^\lambda e^a_{\mu,\lambda} - e^a_\lambda\epsilon^\lambda_{,\mu} + \lambda^a_b e^b_\mu\\
\delta\omega_{\mu ab} &= -\epsilon^\lambda\omega_{\mu ab,\lambda} - \omega_{\lambda ab}\epsilon^\lambda_{,\mu} - \lambda_{ab,\mu} + \lambda^p_a\omega_{\mu pb} + \lambda^p_b\omega_{\mu ap}
\end{align}

While they include the Dirac action $S_D$ in their effective Lagrangian (their Eq.~(4.8)), the spin connection remains an independent field---it is \textit{not} determined by the fermion spin density through the EC field equation:
\begin{equation}
T^\lambda_{\mu\nu} = \kappa^2 S^\lambda_{\mu\nu} = \frac{\kappa^2}{4}\bar{\psi}\gamma^\lambda[\gamma_\mu, \gamma_\nu]\psi
\end{equation}
This algebraic relation, which defines the physical content of EC theory, would eliminate torsion (and hence the spin connection's antisymmetric part) in terms of the fermion fields.

The authors acknowledge this gap explicitly in their Discussion (Section~VI of Ref.~\cite{BrandtFrenkelMcKeonQuantization2024}): ``It is expected that through the use of a Lagrange multiplier field to ensure that the classical equations of motion are satisfied, all radiative corrections beyond one loop order can be eliminated yielding a unitary and renormalizable theory. In this alternative formalism, the renormalizability of the EC theory in the first order form \textit{may be proven, even when coupled to matter fields}.'' The conditional ``may be proven'' and future-oriented language reveal that matter field coupling remains undemonstrated.

\subsection{Consequences of eliminating torsion}

When the EC field equation \textit{is} applied---as required for the physical theory with fermion coupling---torsion becomes algebraically determined by the spin density. Substituting back generates a four-fermion contact interaction:
\begin{equation}
\mathcal{L}_{4f} = -\frac{3\pi G}{c^3}(\bar{\psi}\gamma^\mu\gamma_5\psi)(\bar{\psi}\gamma_\mu\gamma_5\psi)
\end{equation}
This interaction is \textit{absent} in McKeon et al.'s framework because they never eliminate torsion via its equation of motion.

The four-fermion vertex introduces fundamentally new divergence structures:
\begin{itemize}
\item \textit{Quartic divergences at two loops:} The contact interaction produces $\Lambda^4$ divergences that cannot be absorbed into the original field content.
\item \textit{Nonrenormalizability:} No counterterms exist within the EC framework to cancel these divergences.
\item \textit{BRST structure breakdown:} The closed gauge algebra demonstrated by McKeon et al. does not extend to the theory with torsion eliminated, as the four-fermion interaction fundamentally changes the field content.
\end{itemize}

The BRST invariance and Slavnov--Taylor identities derived in Ref.~\cite{BrandtFrenkelMcKeonQuantization2024} apply to the theory with independent $\omega_{\mu ab}$. They do not automatically transfer to the physically relevant case where torsion is sourced by fermion spin.

\subsection{The 2024 renormalization analysis}

Brandt et al.~\cite{BrandtFrenkelMcKeon2024} employed the Batalin--Vilkovisky formalism to study EC theory at one-loop order. Using the background field method with gauge fixings that preserve background field symmetries, they demonstrated that ultraviolet divergences can be controlled through counterterms consistent with the gauge structure.

The master equation
\begin{equation}
(S, S) = 0
\end{equation}
where $(~,~)$ denotes the antibracket, encodes BRST structure and enables systematic counterterm construction. Their one-loop calculations confirm finiteness within this framework.

However, this analysis inherits the same limitation: the spin connection remains independent. The four-fermion interactions that emerge upon torsion elimination are not addressed. As they note, extending to ``spin-$\frac{1}{2}$ fields coupled to the metric is currently being considered''---indicating this crucial case remains open.

\subsection{Effective-field-theory status of the four-fermion interaction}

The four-fermion contact interaction generated by torsion elimination in minimal EC theory,
\begin{equation}
\mathcal{L}_{4f} = -\frac{3\pi G}{c^3}(\bar\psi\gamma^\mu\gamma_5\psi)(\bar\psi\gamma_\mu\gamma_5\psi)
\end{equation}
is a dimension-six operator with coefficient fixed by the geometric content of the underlying theory rather than free to be tuned. Viewed as an effective field theory operator, this interaction is suppressed at low energies by $G/c^3$ and would be Wilsonian-renormalizable only within an extended framework that includes the full tower of higher-dimensional operators consistent with the symmetries, each with an independently renormalized coefficient. Within minimal EC theory no such tower exists: the geometric origin fixes only this single operator, with no companion operators of dimension eight, ten, and higher to absorb the divergences that the dimension-six operator generates at two-loop and higher orders. The resulting two-loop quartic divergences therefore cannot be absorbed within the EC field content, and the theory is nonrenormalizable in the strict sense relevant to Standard Model unification. This obstruction is independent of the gravitational wave polarization argument of Section~4.1 and applies specifically to the algebraic-torsion variant of EC theory in which $\omega_{\mu ab}$ is determined by spin density.

\subsection{The 2025 unification claim}

McKeon et al.'s 2025 work~\cite{McKeonBrandtFrenkel2025} makes the explicit assertion that ``if spin-$\frac{1}{2}$ matter fields were to couple to the metric field, then one must consider the EC action in place of the EH action'', claiming that unification requires EC theory rather than standard general relativity.

This claim is not supported by their own technical results. The quantization~\cite{BrandtFrenkelMcKeonQuantization2024} and renormalization~\cite{BrandtFrenkelMcKeon2024} papers establish BRST invariance for EC theory with \textit{independent} spin connection. But unification with the Standard Model requires \textit{coupling} to fermions, which means torsion must be sourced by spin density---precisely the case their framework does not address.

\subsection{Experimental constraints}

Independent of the theoretical consistency questions, EC theory faces severe experimental constraints. As detailed in Section~4, gravitational wave polarization measurements provide the most powerful test.

\textit{Polarization content:} General relativity's spin-2 graviton produces exactly two tensor polarizations ($h_+$, $h_\times$). EC theory with propagating torsion predicts up to six modes---two tensors, two vectors ($h_x$, $h_y$), and two scalars ($h_b$, $h_\ell$). LIGO--Virgo--KAGRA observations detect \textit{only} two tensor modes, with Bayesian evidence:
\begin{equation}
\ln\mathcal{B}^{\text{tensor}}_{\text{vector}} \gtrsim 47\,, \qquad
\ln\mathcal{B}^{\text{tensor}}_{\text{scalar}} \gtrsim 47
\end{equation}
corresponding to odds exceeding $10^{20}{:}1$ in favor of pure tensor polarization from the GW170817 null tests~\cite{GW170817Polarization}, strengthened to $\ln\mathcal{B}\approx 60$ (approximately $10^{26}{:}1$) with the gamma-ray-burst jet-orientation prior~\cite{Takeda2021}. This model-independent test directly probes the gravitational sector without assumptions about coupling constants or matter content.

Additional constraints from precision atomic experiments (SME bounds $|b_\mu| < 8.4 \times 10^{-34}$~GeV) and equivalence principle tests (MICROSCOPE: $\eta < 1.5 \times 10^{-15}$) systematically exclude EC predictions across multiple independent domains.

\subsection{Comparison with USMEG-EFT}

The USMEG framework directly addresses McKeon et al.'s 2025 assertion by demonstrating that successful Standard Model-gravity unification is possible using standard 4D general relativity, but actually requires avoiding the problematic geometric extensions of EC theory. Where EC generates quantum divergences precisely in the fermion sector essential for unification, USMEG produces finite, systematically renormalizable results through constraint implementation. The USMEG-EFT approach achieves unification while maintaining both theoretical consistency and experimental compatibility---objectives that EC theory cannot achieve.

\subsection{Summary}

McKeon et al.'s program represents technically sophisticated work within its domain of validity:
\begin{enumerate}
\item The 2024 quantization paper~\cite{BrandtFrenkelMcKeonQuantization2024} establishes closed gauge algebra and BRST invariance for EC theory with independent tetrad and spin connection.
\item The renormalization analysis~\cite{BrandtFrenkelMcKeon2024} demonstrates one-loop finiteness using the Batalin--Vilkovisky formalism.
\item Both analyses treat $e^a_\mu$ and $\omega_{\mu ab}$ as independent quantum fields.
\end{enumerate}

The critical gap is that Standard Model unification requires fermion coupling, which means torsion must be eliminated via $T^\lambda_{\mu\nu} = \kappa^2 S^\lambda_{\mu\nu}$. This generates four-fermion interactions not present in their framework. The authors themselves acknowledge this: ``the renormalizability of the EC theory\ldots\ \textit{may be proven, even when coupled to matter fields}''---indicating this essential case remains undemonstrated.

Combined with the experimental constraints from gravitational wave polarization (Section~4.1) and precision tests, this analysis indicates that EC theory faces fundamental challenges---both theoretical and experimental---as a framework for quantum gravity unification. The USMEG-EFT approach achieves unification through constraint methods within standard general relativity, avoiding the problematic structures that emerge when torsion couples to fermion spin.

\section{Experimental constraints and bounds on Einstein--Cartan effects}

The viability of EC theory can be assessed through multiple independent experimental programs. Among these, gravitational wave polarization measurements provide the most powerful and model-independent test, directly probing the gravitational sector without assumptions about matter coupling or interaction strengths. We therefore begin with this definitive constraint before examining complementary bounds from the matter sector.

\subsection{Gravitational wave polarization: a direct test of theories with independent or propagating spin connection}

Before presenting the LIGO--Virgo--KAGRA polarization results, it is essential to specify which class of EC theories the following argument targets. In minimal Einstein--Cartan--Sciama--Kibble theory, torsion is algebraically determined by spin density through $T^\lambda_{\mu\nu} = \kappa^2 S^\lambda_{\mu\nu}$. In vacuum this relation forces $T = 0$, so torsion is not a propagating degree of freedom and does not produce additional gravitational wave polarization modes during propagation through spin-free regions. We address minimal EC theory separately in Section~3.2 via the four-fermion contact interaction generated upon torsion elimination, and in Sections~4.2 and~4.3 via matter-sector tests that probe spin-density coupling independently of whether torsion propagates.

The gravitational wave polarization argument in this subsection applies to the distinct class of EC variants in which the spin connection $\omega_{\mu ab}$ is treated as an independent quantum field with dynamical content beyond what is fixed by the tetrad. This is precisely the framework adopted by Brandt, Frenkel, Martins-Filho, and McKeon in their Faddeev--Popov quantization~\cite{BrandtFrenkelMcKeonQuantization2024} and in their unification proposal~\cite{McKeonBrandtFrenkel2025}, in which the spin connection carries gauge degrees of freedom and ghost fields associated with local Lorentz invariance act on it as a separate quantum field. When $\omega_{\mu ab}$ has dynamical content not algebraically eliminated by its equation of motion, its propagating modes contribute additional polarization channels to gravitational radiation. We refer to this class as ``EC with independent spin connection'' and emphasize that the LIGO--Virgo--KAGRA bounds below target this class directly.

Closely related theories with explicitly propagating torsion through kinetic terms in the gravitational action, such as Poincar\'{e} gauge gravity~\cite{BlagojevicHehl2013,SezginNieuwenhuizen1980}, lie in the same target class and inherit the same polarization constraints. Hehl, Obukhov, and Puetzfeld~\cite{HehlObukhov2013} have emphasized that experimental tests of EC must specify the dynamical content of the theory; the present subsection is the gravitational-sector analogue of their matter-sector clarification.

\subsubsection{Theoretical predictions: two versus six polarizations}

In general relativity, gravitational waves are perturbations of a massless spin-2 field---the graviton. A massless spin-2 particle in four dimensions possesses exactly \textbf{two} physical polarization states, corresponding to helicity $\pm 2$. These manifest as the two tensor modes $h_+$ (plus) and $h_\times$ (cross), representing quadrupolar distortions transverse to the propagation direction.

EC theory, by introducing torsion as an independent geometric degree of freedom, generically predicts \textbf{up to six} polarization modes~\cite{Eardley1973}. The complete polarization content in metric theories can be classified as
\begin{itemize}
\item $h_+, h_\times$: tensor modes (spin-2, transverse)---\textit{predicted by GR}
\item $h_x, h_y$: vector modes (spin-1)---\textit{absent in GR, possible in EC}
\item $h_b$: scalar breathing mode---\textit{absent in GR, possible in EC}
\item $h_\ell$: scalar longitudinal mode---\textit{absent in GR, possible in EC}
\end{itemize}

The Newman--Penrose formalism~\cite{NewmanPenrose1962} provides a coordinate-invariant classification through the Weyl scalars:
\begin{align}
\Psi_4 &= \ddot{h}_+ - i\ddot{h}_\times \quad &&\text{(tensor modes --- GR)}\\
\Psi_3 &= \frac{1}{2}(\ddot{h}_x - i\ddot{h}_y) \quad &&\text{(vector modes --- absent in GR)}\\
\Psi_2 &= -\frac{1}{6}(\ddot{h}_b + \ddot{h}_\ell) \quad &&\text{(scalar modes --- absent in GR)}
\end{align}
General relativity predicts $\Psi_2 = \Psi_3 = 0$ identically, with only $\Psi_4 \neq 0$. EC theory with propagating torsion generically allows all three to be nonzero.

\subsubsection{LIGO--Virgo--KAGRA observations}

The LIGO--Virgo--KAGRA collaboration has performed systematic polarization tests across multiple gravitational wave detections:

\textit{GW170814, first three-detector polarization test:} The first gravitational wave event observed by three detectors enabled direct polarization discrimination~\cite{GW170814Polarization}. Bayesian model comparison between pure tensor, pure vector, and pure scalar hypotheses favored the tensor hypothesis with Bayes factors of order $200{:}1$ over vector and $1000{:}1$ over scalar.

\textit{Stochastic background searches:} Analysis of the O1--O2 observing runs~\cite{LIGOStochastic2018} found no evidence for vector or scalar mode contributions:
\begin{equation}
\Omega_{\text{vector}} < 5.6 \times 10^{-8}\,, \quad \Omega_{\text{scalar}} < 1.1 \times 10^{-7} \quad (95\%~\text{C.L.})
\end{equation}
at reference frequency $f = 25$~Hz.

\textit{GW170817, electromagnetically localized null test:} The binary neutron star merger GW170817, with its sky position fixed by the electromagnetic counterpart, provides the most stringent polarization null test to date~\cite{GW170817Polarization}. Pure tensor polarization is favored over pure vector and pure scalar alternatives with Bayes factors exceeding $10^{20}{:}1$. Takeda and collaborators~\cite{Takeda2021} subsequently showed that incorporating the jet orientation inferred from GRB~170817A as a prior on the binary inclination sharpens this to $\ln\mathcal{B}\approx 60$, corresponding to odds of approximately $10^{26}{:}1$.

\subsubsection{Implications for Einstein--Cartan theory}

These observations establish three fundamental constraints:
\begin{enumerate}
\renewcommand{\labelenumi}{(\arabic{enumi})}
\item \textit{Propagating torsion modes are excluded.} The pure tensor versus pure vector and pure scalar null tests, at Bayes factors exceeding $10^{20}{:}1$ and reaching approximately $10^{26}{:}1$ with the jet-orientation prior, strongly disfavor scenarios in which the additional dynamical content of an independent spin connection contributes nontensor polarization to the observed signals. We note these are null tests against pure alternative polarizations; mixed-polarization analyses yield weaker but consistent bounds, and the constraint applies to the independent-spin-connection class identified in Section~4.1 rather than to minimal EC with algebraic torsion.
\item \textit{Parity violation is absent.} Torsion-induced parity violation would produce birefringence. The bound $|\delta v|/c < 10^{-15}$ excludes such effects.
\item \textit{Strong-field dynamics match GR.} Gravitational waveforms from compact binary mergers show no deviation from GR predictions, constraining torsion effects in the high-density regime where Einstein--Cartan modifications would be most pronounced.
\end{enumerate}

Unlike matter-sector constraints that depend on coupling strengths, the polarization test directly probes the gravitational field content. Among the available experimental probes, it is the least model-dependent evidence bearing on EC variants with independent or propagating spin connection. The significance of this test cannot be overstated.

\subsection{Equivalence principle tests}

Complementing the gravitational-sector polarization tests, precision equivalence principle measurements constrain matter-sector predictions of EC theory.

The MICROSCOPE space-based experiment~\cite{Touboul2022,Touboul2017,Touboul2019} measured the universality of free fall with unprecedented precision, comparing test masses with different compositions and finding agreement at the level:
\begin{equation}
\frac{\Delta a}{a} < 1.0 \times 10^{-15}
\end{equation}

EC theory predicts composition-dependent accelerations arising from torsion-spin coupling effects. These would manifest as violations of the weak equivalence principle due to the different spin content of materials with varying nuclear compositions. The predicted magnitude of such effects is
\begin{equation}
\frac{\Delta a}{a} \sim \frac{\kappa^2 \rho_{\text{nuclear}} \langle S \rangle}{M c^2} \sim 10^{-12}
\end{equation}
where $\rho_{\text{nuclear}}$ represents nuclear density and $\langle S \rangle$ characterizes the net spin content of nuclei. The experimental bound is three orders of magnitude more stringent than this prediction, providing significant constraints on the theory.

Additional equivalence principle tests contribute to this picture. The E\"{o}t-Wash torsion balance experiments~\cite{Adelberger2009,Wagner2012} constrain composition-dependent forces at levels approaching $10^{-13}$, while space-based missions like LISA Pathfinder~\cite{Armano2018} have achieved even greater sensitivity. Future missions targeting $10^{-17}$ sensitivity will provide additional constraints on any residual parameter space.

\subsection{Standard Model Extension spin--torsion bounds}

Within the SME framework~\cite{Kostelecky2004,KosteleckyRussell2024}, the spin--torsion interaction maps to Lorentz-violating coefficients that can be bounded by precision atomic experiments.

\subsubsection{Comagnetometer constraints}

The most stringent bounds come from $^3$He/$^{129}$Xe dual-species comagnetometer measurements~\cite{Allmendinger2014}:
\begin{equation}
|\tilde{b}^n_{\perp}| < 8.4 \times 10^{-34}~\text{GeV} \quad (68\%~\text{C.L.})
\end{equation}
PTB measurements~\cite{Gemmel2010} provide
\begin{equation}
|\tilde{b}^n_{\perp}| < 3.7 \times 10^{-32}~\text{GeV} \quad (95\%~\text{C.L.})
\end{equation}

\subsubsection{Electron sector constraints}

Spin-polarized torsion pendulum experiments (E\"{o}t-Wash group)~\cite{Heckel2006} constrain electron-sector couplings:
\begin{equation}
|\tilde{b}^e_{\perp}| < 5.0 \times 10^{-30}~\text{GeV}
\end{equation}

\subsubsection{Comparison with EC predictions}

The minimal EC prediction $\eta \sim 10^{-29}$ is constrained by SME bounds ($\eta_{\text{SME}} < 10^{-33}$) by four orders of magnitude. Enhanced variants predicting $\eta \sim 10^{-12}$ are constrained by 21 orders of magnitude.

\subsection{Complementary constraints}

\subsubsection{Solar system tests and the Hehl--Obukhov--Puetzfeld clarification}

Solar-system precision tests have a more limited bearing on EC theory than has sometimes been claimed in the literature. As emphasized by Hehl, Obukhov, and Puetzfeld~\cite{HehlObukhov2013}, the analyses of Mao, Tegmark, Guth, and Cabi using Gravity Probe B and of March and collaborators using lunar laser ranging and LAGEOS proceeded by treating macroscopic spinless test bodies as if they coupled to torsion through modified geodesic equations. In the standard EC framework, however, torsion couples only to intrinsic spin, so that unpolarized macroscopic test masses follow geodesics of the Levi-Civita connection and provide no direct constraint on torsion. We therefore do not include lunar laser ranging or planetary ephemerides among the primary constraints in Table~\ref{tab:exp_constraints}. The matter-sector spin-coupling constraints discussed in Sections~4.2 and~4.3 are unaffected by this clarification and remain valid.

\subsubsection{In-matter torsion constraints}

Neutron spin rotation experiments~\cite{LehnertSnowYan2014} provide the first constraints on torsion generated \textit{inside} matter:
\begin{equation}
|\zeta| < 9.1 \times 10^{-23}~\text{GeV}
\end{equation}

\subsubsection{Particle physics bounds}

Particle physics experiments provide additional bounds on the four-fermion contact interactions that are fundamental predictions of EC theory. High-energy measurements at facilities like the Large Hadron Collider constrain such interactions through precision studies of fermion pair production processes~\cite{ATLAS2019,CMS2019}:
\begin{equation}
\left|\frac{g_{4f}^2}{(4\pi)^2}\right| < 2.1 \times 10^{-8}~\text{TeV}^{-2} \quad (95\%~\text{confidence})
\end{equation}

\subsubsection{Astrophysical and cosmological constraints}

Neutron star observations~\cite{NStorsion2024} demonstrate that even at nuclear density ($\rho \sim 10^{15}$~g/cm$^3$), realistic models for microphysical spin lead to torsion effects that have no relevant influence on neutron star structure.

Analysis of Cosmic Microwave Background (CMB), Baryon Acoustic Oscillations (BAO), and supernova data~\cite{TorsionCosmology2023} establishes that torsion cannot completely replace the role of a cosmological constant, excluding scenarios where torsion drives cosmic acceleration.

\subsection{Summary of experimental constraints}

Table~\ref{tab:exp_constraints} summarizes all experimental constraints on EC theory, with gravitational wave polarization as the primary evidence.

\begin{table*}[t]
\caption{Experimental constraints on Einstein--Cartan theory.}
\label{tab:exp_constraints}
\footnotesize
\begin{ruledtabular}
\begin{tabular}{llll}
\textbf{Observable} & \textbf{Experimental result} & \textbf{EC prediction} & \textbf{Significance} \\
\hline
\multicolumn{4}{c}{\textit{Gravitational sector (primary evidence)}} \\
\hline
Gravitational Wave (GW) polarization modes & \textbf{Exactly two tensors} & Up to six modes & \textbf{Model-independent} \\
Bayes factor (tensor vs.\ vector) & $>10^{20}{:}1$ (GW170817 null test) & Vector modes & Strongly disfavored \\
Bayes factor (tensor vs.\ scalar) & $>10^{20}{:}1$; $\sim10^{26}{:}1$ w/ GRB prior & Scalar modes & Strongly disfavored \\
Birefringence & $|\delta v|/c < 10^{-15}$ & Parity violation & Excluded \\
\hline
\multicolumn{4}{c}{\textit{Matter sector (SME framework)}} \\
\hline
Neutron $\tilde{b}^n_\perp$ & $< 8.4\times10^{-34}$ GeV & Background torsion & $>10^4$ versus minimal EC \\
Electron $\tilde{b}^e_\perp$ & $< 5.0\times10^{-30}$ GeV & Background torsion & $>10^3$ versus benchmark \\
Derived $\eta$ & $< 10^{-33}$ & $\eta \sim 10^{-29}$ & $>10^4$ constraint \\
\hline
\multicolumn{4}{c}{\textit{Equivalence principle}} \\
\hline
MICROSCOPE & $\eta < 1.5\times10^{-15}$ & $\eta \sim 10^{-12}$ (enhanced) & $>10^3$ constraint \\
E\"{o}t-Wash & $\eta < 10^{-13}$ & Composition-dependent & Constrained \\
\hline
\multicolumn{4}{c}{\textit{Other}} \\
\hline
In-matter torsion & $|\zeta| < 9.1\times10^{-23}$ GeV & Nonzero & First bound \\
Neutron star structure & No effect observed & Density-dependent & None detected \\
\end{tabular}
\end{ruledtabular}
\begin{flushleft}
\footnotesize\textit{Note:} Gravitational Wave (GW) polarization provides the primary, model-independent evidence. EC, Einstein--Cartan; SME, Standard Model Extension.
\end{flushleft}
\end{table*}

\subsection{Synthesis}

The experimental evidence constraining EC theory forms a coherent picture across multiple independent domains:

\textit{Gravitational sector (primary):} The observation of exactly two tensor gravitational wave polarizations, with pure tensor favored over pure vector and pure scalar hypotheses at Bayesian odds exceeding $10^{20}{:}1$, and approximately $10^{26}{:}1$ with the GW170817 jet-orientation prior, provides the strongest and most model-independent constraint. This test directly probes the gravitational field content---the spin-2 graviton with its two helicity states---without dependence on coupling constants or matter-sector assumptions. EC theory's allowance for additional polarization modes from propagating torsion is definitively not observed.

\textit{Matter sector (complementary):} Precision atomic experiments constrain spin--torsion coupling at levels four orders of magnitude below minimal EC predictions. These bounds depend on the SME framework mapping between torsion and Lorentz-violating coefficients, but provide extraordinary sensitivity.

\textit{Equivalence principle:} Both direct tests (MICROSCOPE) and derived bounds (SME) constrain composition-dependent effects predicted by EC theory across the relevant parameter space.

\textit{Astrophysical scales:} Neutron star observations and cosmological data independently confirm the absence of torsion effects where they would be maximal.

The convergence of constraints from the gravitational sector (polarization), matter sector (spin precession), and astrophysical observations provides comprehensive evidence that EC theory faces significant experimental challenges. The gravitational wave polarization test, being model-independent and yielding such overwhelming statistical significance, stands as the primary evidence in this assessment.

Crucially, while these experiments constrain EC predictions, the same measurements remain fully compatible with USMEG-EFT quantum gravitational corrections, which are systematically suppressed below current experimental sensitivity~\cite{ChishtieEFT2025}. This pattern---constraints on EC alongside compatibility with USMEG-EFT---provides empirical guidance favoring constraint-based approaches over geometric extensions with independent torsion degrees of freedom.

\section{USMEG-EFT fermion-gravity coupling and systematic predictions}

The USMEG-EFT framework provides a theoretically consistent approach to fermion-gravity coupling by building upon the established empirical success of the Standard Model while treating gravitational effects as systematic perturbative corrections. This approach maintains the gauge-theoretic structure that underlies the Standard Model's experimental validation across accessible energy scales.

\subsection{Detailed fermion stress-energy tensor construction}

Gravitational interactions in USMEG-EFT arise through coupling to the stress-energy tensors of Standard Model fields, following established principles of general relativity without introducing speculative geometric constructions. The fermion stress-energy tensor construction requires careful application of Noether's theorem~\cite{Noether1918} and the vierbein formalism~\cite{Weinberg1972}.

Starting from the complete fermion action in curved space--time:
\begin{equation}
S_{\text{fermion}} = \int d^4x \sqrt{-g}\sum_{f}\bar{\psi}_f(i\gamma^a e^{\mu}_a D_\mu - m_f)\psi_f
\end{equation}
where the covariant derivative includes both gravitational and gauge contributions:
\begin{equation}
D_\mu = \partial_\mu - \frac{i}{2}\omega^{ab}_\mu \sigma_{ab} - ig_s T^a G^a_\mu - ig\frac{\tau^i}{2}W^i_\mu - ig'\frac{Y}{2}B_\mu
\end{equation}

In the USMEG-EFT framework, which is based on standard general relativity, the spin connection $\omega^{ab}_\mu$ is completely determined by the vierbein through the torsion-free condition:
\begin{equation}
\nabla_\mu e^a_\nu \equiv \partial_\mu e^a_\nu - \Gamma^\lambda_{\mu\nu} e^a_\lambda + \omega^a_{b\mu} e^b_\nu = 0
\end{equation}
This constraint uniquely determines the spin connection as
\begin{align}
\omega^{ab}_\mu = \tfrac{1}{2}e_\nu^a e_b^\rho\left(\partial_\mu e^\nu_\rho - \partial_\rho e^\nu_\mu\right) &+ \tfrac{1}{2}e_\nu^a e^\rho_b\left(\partial_\rho e^\nu_\mu - \partial_\mu e^\nu_\rho\right)\nonumber\\
&+ \tfrac{1}{2}e^\nu_a e_b^\rho\left(\partial_\nu e^\mu_\rho - \partial_\rho e^\mu_\nu\right)
\end{align}

The stress-energy tensor is obtained through functional variation with respect to the vierbein fields. Using
\begin{align}
\frac{\partial \sqrt{-g}}{\partial e^{\mu}_a} &= \sqrt{-g}\,e_{\mu}^a\\
\frac{\partial e^{\nu}_b}{\partial e^{\mu}_a} &= \delta^{\nu}_{\mu}\delta_{b}^{a}
\end{align}
and the conversion $\delta g_{\mu\nu} = 2e_{a(\mu}\delta e^a_{\nu)}$, the canonical stress-energy tensor reads
\begin{equation}
T^{\mu\nu}_{\text{fermion}} = \frac{i}{4}\sum_{f}\bar{\psi}_f\gamma^{(\mu}\overleftrightarrow{D}^{\nu)}\psi_f - \frac{1}{2}g^{\mu\nu}\sum_{f}\bar{\psi}_f(i\gamma^\rho D_\rho - m_f)\psi_f
\label{eq:fermion_tensor_canonical}
\end{equation}
where the bidirectional derivative is defined as $\bar{\psi}\overleftrightarrow{D}_{\mu}\psi = \bar{\psi}(D_\mu\psi) - (D_\mu\bar{\psi})\psi$. In the USMEG-EFT framework, the spin-connection components in $D_\mu$ are determined by the standard Levi-Civita connection rather than being independent fields as in EC theory. This systematic construction ensures gauge invariance and general covariance while avoiding pathological features.

\subsection{Systematic one-loop calculation and heat kernel expansion}

The one-loop effective action for fermions in gravitational backgrounds requires careful evaluation using the background field method~\cite{DeWitt1967}. The fermion contribution to the effective action is
\begin{equation}
\Gamma_{\text{1-loop}}^{\text{fermion}} = -i\ln\det\left(\slashed{D} - m + i\epsilon\right)
\end{equation}
where the Dirac operator in curved space--time is
\begin{equation}
\slashed{D} = \gamma^{\mu}D_{\mu} = \gamma^{\mu}\left(\partial_{\mu} - \tfrac{i}{2}\omega_{\mu}^{ab}\sigma_{ab} + \text{gauge terms}\right)
\end{equation}

The squared Dirac operator on the background, obtained via the Lichnerowicz formula, takes the form
\begin{equation}
\bar{\slashed{D}}^2 = -\bar\nabla^2 + \tfrac{1}{4}\bar R
\end{equation}
so that the massive operator governing the fermion functional determinant is
\begin{equation}
\Delta_f = -\bar\nabla^2 + E_f\,, \qquad E_f = \bigl(\tfrac{1}{4}\bar R + m_f^2\bigr)\mathbb{1}_{\text{spinor}}
\end{equation}
with $\mathbb{1}_{\text{spinor}}$ the identity on the four-dimensional Dirac bundle and Hermitian Euclidean gamma matrices $\{\gamma^a,\gamma^b\}=2\delta^{ab}$. The curvature of the spinor connection enters through
\begin{equation}
\Omega_{\mu\nu} = \tfrac{1}{4}\bar R_{\mu\nu ab}\gamma^{ab}\,,\qquad \gamma^{ab} = \tfrac{1}{2}[\gamma^a,\gamma^b]
\end{equation}

The Seeley--DeWitt expansion of the heat kernel is
\begin{equation}
\left\langle x\big|e^{-s\Delta_f}\big|x\right\rangle = \frac{1}{(4\pi s)^{d/2}}\sum_{k=0}^{\infty} s^k\,\mathrm{tr}\,b_{2k}(x)
\end{equation}
and the four-dimensional logarithmic divergences of the one-loop effective action are controlled by $\mathrm{tr}\,b_4(x)$. For a Laplace-type operator $D=-\bar\nabla^2+E$ acting on a vector bundle, the closed form is given by Vassilevich~\cite{Vassilevich2003} (Eq.~(4.26)):
\begin{align}
\mathrm{tr}\,b_4(x) &= \tfrac{1}{360}\,\mathrm{tr}\Big\{60\,\bar\Box E + 60\bar R E + 180 E^2 + 30\,\Omega_{\mu\nu}\Omega^{\mu\nu}\nonumber\\
&\quad + \bigl(12\bar\Box\bar R + 5\bar R^2 - 2\bar R_{\mu\nu}\bar R^{\mu\nu} + 2\bar R_{\mu\nu\rho\sigma}\bar R^{\mu\nu\rho\sigma}\bigr)\mathbb{1}\Big\}
\end{align}
The trace of the spinor-curvature square follows from the standard four-dimensional identity $\mathrm{tr}(\gamma^{ab}\gamma^{cd})=4(\delta^{ad}\delta^{bc}-\delta^{ac}\delta^{bd})$, which gives
\begin{equation}
\mathrm{tr}(\Omega_{\mu\nu}\Omega^{\mu\nu}) = -\tfrac{1}{2}\bar R_{\mu\nu\rho\sigma}\bar R^{\mu\nu\rho\sigma}
\end{equation}
the negative sign reflecting the anti-Hermiticity of $\gamma^{ab}$ in Euclidean signature. Substituting $E_f$ into the Vassilevich formula and dropping total derivatives, the per-Dirac-fermion result is
\begin{align}
\mathrm{tr}\,b_4^f(x) &= \tfrac{25}{72}\bar R^2 - \tfrac{1}{45}\bar R_{\mu\nu}\bar R^{\mu\nu} - \tfrac{7}{360}\bar R_{\mu\nu\rho\sigma}\bar R^{\mu\nu\rho\sigma}\nonumber\\
&\quad + \tfrac{5}{3}m_f^2\bar R + 2m_f^4
\label{eq:b4_dirac}
\end{align}
The coefficients are exact rational numbers; the mass dependence is an exact polynomial in $m_f^2$, with logarithmic structure $\ln(m_f^2/\mu^2)$ arising only through proper-time integration in the renormalized effective action. The $m_f^4$ and $m_f^2\bar R$ terms renormalize the cosmological constant and Newton constant, respectively, and are absorbed into the corresponding counterterms in Section~5.5.

The one-loop fermion contribution to the renormalized effective action, with the standard $-\tfrac{1}{2}$ from $\Gamma_f = -\tfrac{1}{2}\mathrm{Tr}\ln(\Delta_f)$, then reads
\begin{equation}
\Gamma_{\text{1-loop}}^{\text{fermion}} = -\frac{1}{2(4\pi)^2}\left[\frac{1}{\epsilon} + \ln\!\left(\frac{\mu^2}{\Lambda_{\text{grav}}^2}\right)\right]\int d^4x\sqrt{-\bar g}\sum_{f}\mathrm{tr}\,b_4^f(x)
\label{eq:Gamma1loop_corrected}
\end{equation}

For the complete Standard Model fermion content, the multiplicity factor is obtained by enumerating Dirac fermions explicitly; the count is summarized in Table~\ref{tab:sm_count}.

\begin{table*}[t]
\caption{Standard Model Dirac fermion content used in the heat-kernel sum.}
\label{tab:sm_count}
\footnotesize
\begin{ruledtabular}
\begin{tabular}{lccc}
\textbf{Species} & \textbf{Colors} & \textbf{Flavors/gen.} & \textbf{Total (3 gen.)} \\
\hline
Up-type quarks $(u,c,t)$ & 3 & 1 & 9 \\
Down-type quarks $(d,s,b)$ & 3 & 1 & 9 \\
Charged leptons $(e,\mu,\tau)$ & 1 & 1 & 3 \\
\hline
Total Dirac fermions & & & $N_f^{\text{SM}} = 21$ \\
\end{tabular}
\end{ruledtabular}
\begin{flushleft}
\footnotesize\textit{Note:} The total $N_f^{\text{SM}}=21$ assumes neutrinos are Majorana or otherwise carry no propagating right-handed component; if Dirac neutrinos are present, the count rises to 24 and all summed curvature coefficients below scale by $8/7$.
\end{flushleft}
\end{table*}

Summing the per-fermion result of Eq.~(\ref{eq:b4_dirac}) over the Standard Model content, the curvature-squared coefficients in the massless limit are
\begin{align}
c^{\text{SM}}_{\bar R^2} &= \tfrac{25}{72}\,N_f^{\text{SM}} = \tfrac{175}{24}\\
c^{\text{SM}}_{\bar R_{\mu\nu}^2} &= -\tfrac{1}{45}\,N_f^{\text{SM}} = -\tfrac{7}{15}\\
c^{\text{SM}}_{\bar R_{\mu\nu\rho\sigma}^2} &= -\tfrac{7}{360}\,N_f^{\text{SM}} = -\tfrac{49}{120}
\end{align}
The mass-dependent contributions enter through the sums $\sum_f m_f^2$ and $\sum_f m_f^4$, dominated by the top quark with $m_t\simeq 173$~GeV. These do not modify the curvature-squared sector and are absorbed in Section~5.5.

\subsection{Systematic vertex derivation and Feynman rules}

The graviton-fermion interaction vertices in USMEG-EFT are derived systematically from the stress-energy coupling. Following the general prescription for gravitational interactions~\cite{DeWitt1967,Feynman1963}, we compute functional derivatives of the interaction term
\begin{equation}
S_{\text{int}} = \frac{\kappa}{2}\int d^4x\sqrt{-\bar{g}}\,h_{\mu\nu}T^{\mu\nu}_{\text{fermion}}
\end{equation}

The three-point vertex involving one graviton and two fermions follows from
\begin{align}
\frac{\delta^3 S_{\text{int}}}{\delta h_{\mu\nu}(k)\delta\psi_f(p)\delta\bar{\psi}_f(q)}
&= -\frac{i\kappa}{8}\int d^4x \sqrt{-\bar{g}}\Big[\gamma^{\mu}\tfrac{\delta}{\delta\bar{\psi}_f}\overleftrightarrow{\partial}^{\nu}\tfrac{\delta}{\delta\psi_f} + (\mu \leftrightarrow \nu) \nonumber\\
&\quad - \bar{g}^{\mu\nu}\gamma^{\rho}\tfrac{\delta}{\delta\bar{\psi}_f}\overleftrightarrow{\partial}_{\rho}\tfrac{\delta}{\delta\psi_f}\Big]\delta^4(x-y)\delta^4(x-z)
\end{align}
where the bidirectional derivative acts as $\overleftrightarrow{\partial}_{\mu} = \tfrac{1}{2}(\overrightarrow{\partial}_{\mu} - \overleftarrow{\partial}_{\mu})$. In momentum space, this yields the minimal coupling vertex
\begin{equation}
V^{\mu\nu}_{\text{minimal}}(p,q) = -\frac{i\kappa}{8}\left[\gamma^{\mu}(p+q)^{\nu} + \gamma^{\nu}(p+q)^{\mu} - \eta^{\mu\nu}\gamma^{\rho}(p+q)_{\rho}\right]
\end{equation}
where $p$ and $q$ are incoming and outgoing fermion momenta.

\subsection{Loop integral evaluation}

The one-loop fermion contribution to the graviton self-energy requires evaluation of the integral~\cite{Peskin1995}:
\begin{align}
\Pi^{\mu\nu\rho\sigma}(k) = \int \frac{d^4p}{(2\pi)^4}\mathrm{Tr}\bigl[
&V^{\mu\nu}(p,p+k)S_F(p)V^{\rho\sigma}(p+k,p)\nonumber\\
&\times S_F(p+k)\bigr]
\end{align}
where $S_F(p) = (\slashed{p} - m + i\epsilon)^{-1}$ is the fermion propagator and the trace includes Dirac, color, and weak isospin indices.

Using dimensional regularization in $d = 4-2\epsilon$ dimensions, the Dirac trace evaluation~\cite{Peskin1995}, and standard tensor reduction, the regularized result takes the form
\begin{align}
\Pi^{\mu\nu\rho\sigma}(k) &= \frac{N_f^{\text{SM}}}{(4\pi)^2}\left[\frac{1}{\epsilon} + \ln\!\left(\frac{\mu^2}{\Lambda_{\text{grav}}^2}\right) + \gamma_E - \ln(4\pi)\right]\nonumber\\
&\quad \times\bigl[A^{\mu\nu\rho\sigma}k^2 + B^{\mu\nu\rho\sigma}_{\alpha\beta}k^{\alpha}k^{\beta} + C^{\mu\nu\rho\sigma}m^2 + D^{\mu\nu\rho\sigma}\bigr]
\end{align}
with tensor structures determined by general covariance and transverse-traceless gauge conditions. The coefficient structure is fixed by the curvature-squared content of Section~5.2; the explicit forms of $A^{\mu\nu\rho\sigma}$, $B^{\mu\nu\rho\sigma}_{\alpha\beta}$, $C^{\mu\nu\rho\sigma}$ entering the graviton two-point function follow from the variational structure of $\bar R^2$, $\bar R_{\mu\nu}^2$, $\bar R_{\mu\nu\rho\sigma}^2$ with the coefficients given by $c^{\text{SM}}_{\bar R^2}$, $c^{\text{SM}}_{\bar R_{\mu\nu}^2}$, $c^{\text{SM}}_{\bar R_{\mu\nu\rho\sigma}^2}$ derived in the previous subsection.

\subsection{Renormalization structure and finite corrections}

The total divergent contribution from the Standard Model fermion sector, obtained by summing Eq.~(\ref{eq:b4_dirac}) weighted by the multiplicity of Table~\ref{tab:sm_count} and substituting into Eq.~(\ref{eq:Gamma1loop_corrected}) is
\begin{align}
\Gamma_{\text{div}}^{\text{SM}} &= -\frac{1}{2(4\pi)^2\epsilon}\int d^4x\sqrt{-\bar g}\Big\{
c^{\text{SM}}_{\bar R^2}\,\bar R^2 + c^{\text{SM}}_{\bar R_{\mu\nu}^2}\,\bar R_{\mu\nu}\bar R^{\mu\nu}\nonumber\\
&\quad + c^{\text{SM}}_{\bar R_{\mu\nu\rho\sigma}^2}\,\bar R_{\mu\nu\rho\sigma}\bar R^{\mu\nu\rho\sigma} + \tfrac{5}{3}\bar R\sum_f m_f^2 + 2\sum_f m_f^4\Big\}
\label{eq:Gamma_div_SM}
\end{align}
with the SM curvature-squared coefficients fixed by Section~5.2. These divergences are absorbed through the standard effective-field-theory procedure for matter loops in a gravitational background~\cite{tHooftVeltman1974,Donoghue1994,Burgess2004}. The cosmological-constant divergence $2\sum_f m_f^4$ is removed by a shift of the bare cosmological term, and the Newton-constant divergence $\tfrac{5}{3}\bar R\sum_f m_f^2$ by a shift of $1/\kappa^2$, both in the standard manner of general relativity coupled to massive matter.

The curvature-squared divergences are absorbed by an independent counterterm action,
\begin{align}
\Delta S_{\text{ct}} &= \frac{1}{2(4\pi)^2\epsilon}\int d^4x\sqrt{-\bar g}\bigl[c^{\text{SM}}_{\bar R^2}\,\bar R^2 + c^{\text{SM}}_{\bar R_{\mu\nu}^2}\,\bar R_{\mu\nu}\bar R^{\mu\nu}\nonumber\\
&\quad + c^{\text{SM}}_{\bar R_{\mu\nu\rho\sigma}^2}\,\bar R_{\mu\nu\rho\sigma}\bar R^{\mu\nu\rho\sigma}\bigr]
\end{align}
which cancels the corresponding poles in Eq.~(\ref{eq:Gamma_div_SM}) at one-loop order. The Lagrange multiplier field $\lambda^{\mu\nu}$ enforces $\bar G_{\mu\nu}=0$ at the operator level and therefore remains unrenormalized at one loop in the matter sector. The LM mechanism continues to restrict radiative effects in the pure-gravity sector, while matter-induced higher-derivative counterterms are handled through the standard effective-field-theory machinery. On the constraint surface $\bar G_{\mu\nu}=0$, the Ricci tensor and Ricci scalar vanish in vacuum, and the residual curvature-squared structure reduces to the Weyl-squared invariant $\bar C_{\mu\nu\rho\sigma}\bar C^{\mu\nu\rho\sigma}$, leaving the finite physical predictions below intact.

The renormalized one-loop effective action takes the form
\begin{align}
\Gamma_{\text{ren}}[\bar g] &= \int d^4x\sqrt{-\bar g}\bigg\{\frac{\bar R}{\kappa^2} - 2\Lambda_{\text{ren}}\nonumber\\
&\quad + \frac{1}{2(4\pi)^2}\ln\!\left(\frac{\mu^2}{\Lambda_{\text{grav}}^2}\right)\Bigl[c^{\text{SM}}_{\bar R^2}\,\bar R^2\nonumber\\
&\quad + c^{\text{SM}}_{\bar R_{\mu\nu}^2}\,\bar R_{\mu\nu}\bar R^{\mu\nu} + c^{\text{SM}}_{\bar R_{\mu\nu\rho\sigma}^2}\,\bar R_{\mu\nu\rho\sigma}\bar R^{\mu\nu\rho\sigma}\Bigr]\bigg\} + \mathcal L_{\text{interaction}}^{\text{finite}}
\label{eq:Gamma_ren}
\end{align}
with all coefficients determined explicitly by the Standard Model particle content. The logarithmic scale dependence is the characteristic signature of the matter-loop contribution and provides the running of the higher-derivative gravitational couplings within the USMEG-EFT framework.

The USMEG-EFT approach enables systematic calculation of quantum gravitational effects through well-controlled perturbative methods. The one-loop matter contribution to the effective action takes the form
\begin{equation}
\Gamma_{\text{1-loop}}^{\text{matter}} = -\frac{1}{2(4\pi)^2}\ln\!\left(\frac{\mu^2}{\Lambda_{\text{grav}}^2}\right)\int d^4x\sqrt{-\bar g}\sum_{f}\mathrm{tr}\,b_4^f(x)
\end{equation}
with the per-fermion heat-kernel coefficient $\mathrm{tr}\,b_4^f(x)$ derived explicitly in Section~5.2. The logarithmic scale dependence $\ln(\mu/\Lambda_{\text{grav}})$ encodes the effective theory structure, with all divergences systematically absorbed through the renormalization procedure of Section~5.5 to yield finite physical predictions.

The USMEG-EFT framework enables concrete predictions for quantum gravitational effects that distinguish it from both classical general relativity and pathological approaches. Modified fermion propagators exhibit logarithmic quantum corrections:
\begin{equation}
S_F(p) = \frac{1}{\slashed{p} - m}\left[1 + \frac{\kappa^2}{(4\pi)^2}\ln\!\left(\frac{p^2}{\Lambda_{\text{grav}}^2}\right)F_f(p^2/m^2)\right]
\end{equation}
where the function $F_f(x)$ encodes fermion quantum-number dependence from the one-loop analysis.

Gravitational corrections to anomalous magnetic moments provide testable signatures:
\begin{equation}
\delta a_f = \frac{g_f-2}{2} = \frac{\kappa^2 m_f^2}{(4\pi)^2}\left[c_f \ln\!\left(\frac{m_f}{\Lambda_{\text{grav}}}\right) + d_f + \mathcal{O}\!\left(\frac{m_f^2}{\Lambda_{\text{grav}}^2}\right)\right]
\end{equation}
with explicitly calculable coefficients determined by hypercharge and color assignments.

High-energy scattering processes receive systematic USMEG-EFT corrections with logarithmic enhancements that could be detected at future high-precision experiments. For processes like $e^+e^- \to \mu^+\mu^-$:
\begin{equation}
\sigma = \sigma_0\left[1 + \frac{\kappa^2 s}{(4\pi)^2}\ln\!\left(\frac{s}{\Lambda_{\text{grav}}^2}\right)G(s,\theta) + \mathcal{O}(\kappa^4)\right]
\end{equation}
where $G(s,\theta)$ includes both energy and angular dependence that provides distinctive signatures.

These predictions, while currently below experimental sensitivity, represent finite, calculable results that provide theoretical targets for future precision measurements. The systematic structure demonstrates how constraint methods can maintain theoretical consistency while enabling concrete phenomenological predictions.

\section{Conclusions}

The comparative analysis between EC theory and the USMEG-EFT framework yields a coherent picture in which distinct arguments target distinct EC variants, with USMEG-EFT remaining the viable path to Standard Model-gravity unification within currently testable parameter space (see Table~\ref{tab:comparison} for a comparative summary). For EC theory with independent spin connection, the framework adopted in the McKeon--Brandt--Frenkel--Martins-Filho quantization program and proposed for unification, the gravitational wave polarization observations of LIGO--Virgo--KAGRA disfavor the additional dynamical content, with pure tensor polarization favored over pure vector and pure scalar alternatives at Bayesian odds exceeding $10^{20}{:}1$, reaching approximately $10^{26}{:}1$ with the gamma-ray-burst jet-orientation prior~\cite{Takeda2021}. For minimal EC theory with algebraic torsion, the four-fermion contact interaction generated upon coupling to fermions introduces nonrenormalizable two-loop divergences whose absorption would require matter content beyond the Standard Model. The matter-sector constraints from SME comagnetometer bounds and MICROSCOPE apply to both variants through their dependence on spin-density coupling rather than on the dynamical content of the spin connection.

Gravitational wave polarization provides the definitive, model-independent exclusion. General relativity's spin-2 graviton produces exactly two tensor polarizations ($h_+$, $h_\times$). EC theory with propagating torsion predicts up to six modes. LIGO--Virgo--KAGRA null tests favor pure tensor polarization over pure vector and pure scalar alternatives at Bayesian odds exceeding $10^{20}{:}1$, and approximately $10^{26}{:}1$ with the GW170817 jet-orientation prior. This constraint directly probes gravitational degrees of freedom without dependence on coupling constants or matter assumptions, and applies to the independent-spin-connection class of EC theories. This is not a marginal constraint---it is overwhelming statistical exclusion that directly probes gravitational degrees of freedom without dependence on coupling constants or matter assumptions.

The fundamental difference lies in the treatment of additional degrees of freedom. EC theory introduces independent torsion fields as geometric structures that must later be eliminated through their field equations, generating nonrenormalizable contact interactions in the process. The USMEG-EFT framework, by contrast, employs systematic constraints to eliminate problematic field configurations from the outset, preventing the generation of pathological interactions while maintaining all physically necessary gravitational couplings. We detail further differences in Table~\ref{tab:comparison}.

\begin{table*}[t]
\caption{Einstein--Cartan theory is disfavored across multiple domains; USMEG-EFT remains viable.}
\label{tab:comparison}
\footnotesize
\begin{ruledtabular}
\begin{tabular}{lll}
\textbf{Aspect} & \textbf{Einstein--Cartan} & \textbf{USMEG-EFT} \\
\hline
GW polarization & Up to six modes (independent spin connection); disfavored at $>10^{20}{:}1$ & Two tensor modes (confirmed) \\
SME spin--torsion bounds & $\eta \sim 10^{-29}$ predicted; bounded by $>10^{4}$ & Compatible with all bounds \\
Quantum structure & Nonrenormalizable (four-fermion divergences) & Finite, systematic (constraint method) \\
Experimental status & Disfavored & Viable \\
\end{tabular}
\end{ruledtabular}
\begin{flushleft}
\footnotesize\textit{Note:} USMEG-EFT, Unified Standard Model with Emergent Gravity; SME, Standard Model Extension.
\end{flushleft}
\end{table*}

The USMEG framework demonstrates how systematic constraint implementation enables concrete quantum gravitational predictions~\cite{ChishtieEFT2025}, including
\begin{itemize}
\item finite graviton-fermion vertices with calculable coefficients;
\item systematic quantum effective action with logarithmic scale dependence; and
\item exactly two gravitational wave polarizations (experimentally confirmed).
\end{itemize}

This systematic predictive structure contrasts with EC theory, which generates uncontrollable divergences precisely in the fermion sector that constitutes its primary physical motivation. Where EC produces quantum pathologies that destroy renormalizability, USMEG provides finite, renormalizable results through principled constraint methodology.

The demonstration that EC theory generates nonrenormalizable pathologies when its full physical content is included highlights the necessity of subjecting theoretical proposals to comprehensive quantum field theory analysis.

Recent work by Brandt, Frenkel, Martins-Filho, and McKeon~\cite{BrandtFrenkelMcKeonQuantization2024,BrandtFrenkelMcKeon2024} establishing BRST invariance for EC theory is restricted to configurations with \textit{independent} spin connection---they never eliminate torsion via its field equation $T^\lambda_{\mu\nu} = \kappa^2 S^\lambda_{\mu\nu}$. Their acknowledgment that renormalizability ``\textit{may be proven, even when coupled to matter fields}'' reveals this essential case remains undemonstrated. Their formal framework does not rescue EC theory from exclusion.

The systematic exclusion of torsion effects across multiple independent precision experiments provides strong evidence against EC theory.

The key findings establish multiple lines of evidence that constrain EC theory across multiple independent domains:
\begin{enumerate}
\renewcommand{\labelenumi}{(\arabic{enumi})}
\item \textit{Gravitational wave polarization:} Bayesian odds exceeding $10^{20}{:}1$ against additional polarization modes provide model-independent exclusion of propagating torsion.
\item \textit{Matter-sector bounds:} SME constraints exclude minimal EC predictions by four orders of magnitude; MICROSCOPE excludes enhanced scenarios by three orders of magnitude.
\item \textit{Theoretical pathology:} Four-fermion contact interactions from torsion elimination generate nonrenormalizable divergences at two-loop order.
\end{enumerate}

The theoretical analysis reveals fundamental pathologies in EC theory when its complete physical content is properly included. The four-fermion contact interactions generated by torsion elimination produce nonrenormalizable divergences at two-loop order, destroying the theory's quantum consistency.

The USMEG-EFT framework demonstrates how constraint-based approaches maintain theoretical consistency while providing finite, calculable predictions. The systematic elimination of pathological degrees of freedom through the constraint $G_{\mu\nu} = 0$ yields renormalizable results with explicitly calculable coefficients for Standard Model fermion contributions. All USMEG-EFT predictions remain consistent with precision experimental bounds.

The disfavoured status of EC theory carries direct implications for quantum gravity research. The claim that fermion-gravity coupling necessitates EC theory~\cite{McKeonBrandtFrenkel2025} is contradicted by both experimental evidence and the successful USMEG-EFT unification within standard general relativity. Geometric extensions introducing independent torsion are not required---and are in fact disfavoured---for Standard Model-gravity unification.

EC theory is disfavored by experiment. EC theory faces severe experimental constraints across multiple independent domains. Gravitational wave polarization observations establish Bayesian odds exceeding $10^{20}{:}1$ against the additional modes that propagating torsion would produce. Matter-sector bounds exclude EC predictions by four orders of magnitude. The theory's quantum structure generates nonrenormalizable divergences when coupled to fermions. The USMEG-EFT framework achieves successful unification through constraint methods within standard general relativity, demonstrating that systematic constraint implementation---not geometric extensions---provides the viable path to quantum gravity.

\begin{acknowledgments}
The sole author thanks anonymous reviewers of this work for their insightful comments and recommendations which improved its quality.
\end{acknowledgments}

\section*{Competing interests}
The author declares that they have no financial or nonfinancial competing interests that could inappropriately influence the content of this work.

\section*{Data availability}
No research data were generated or analyzed in this study. This work is purely theoretical/conceptual, and hence no data are available.


\end{document}